%% file: main.tex
\documentclass[10pt,letterpaper]{article}

\usepackage[letterpaper,top=0.72in,bottom=0.78in,left=0.72in,right=0.72in,
            columnsep=0.25in]{geometry}
\usepackage[T1]{fontenc}
\usepackage[utf8]{inputenc}
\usepackage{lmodern}
\usepackage{microtype}
\usepackage[hyphens]{url}
\usepackage{graphicx}
\usepackage[round,authoryear]{natbib}
\usepackage[font=small,labelfont=bf]{caption}
\usepackage{booktabs}
\usepackage{array}
\usepackage{amsmath}
\usepackage{amssymb}
\usepackage[table]{xcolor}
\usepackage{listings}
\usepackage{hyperref}
\usepackage{titlesec}

\titleformat{\section}{\large\bfseries}{\thesection}{1em}{}
\titleformat{\subsection}{\normalsize\bfseries}{\thesubsection}{1em}{}
\titleformat{\subsubsection}{\small\bfseries}{\thesubsubsection}{1em}{}

\definecolor{linkblue}{HTML}{1F5A7A}
\definecolor{promptgray}{HTML}{F4F5F6}
\hypersetup{
  colorlinks=true,
  linkcolor=linkblue,
  citecolor=linkblue,
  urlcolor=linkblue,
  pdfauthor={Zackary Okun Dunivin},
  pdftitle={Who Uses Open-Weight Models? China and the Shifting Geography of AI in Science}
}
\providecommand{\sisetup}[1]{}

\lstnewenvironment{Prompt}{%
  \lstset{
    basicstyle=\ttfamily\fontsize{6}{6.8}\selectfont,
    backgroundcolor=\color{promptgray},
    frame=single,
    framerule=0.25pt,
    rulecolor=\color{black!25},
    breaklines=true,
    breakatwhitespace=false,
    columns=fullflexible,
    keepspaces=true,
    showstringspaces=false,
    xleftmargin=0.25em,
    xrightmargin=0.25em,
    aboveskip=0.6em,
    belowskip=0.6em
  }%
}{}

\title{Who Uses Open-Weight Models? China and the Shifting Geography of AI in Science}
\author{Zackary Okun Dunivin\\
\small Institute for Social Sciences, University of Stuttgart\\
\small Stuttgart, Germany\\
\small \href{mailto:zackary.dunivin@sowi.uni-stuttgart.de}{zackary.dunivin@sowi.uni-stuttgart.de}}
\date{}

\begin{document}
\fontsize{10pt}{11.2pt}\selectfont
\twocolumn[{
\maketitle

\begin{abstract}
As LLMs have become a flashpoint for scientific research, computer scientists and STS scholars have advocated the use of open-weight (ostensibly inspectable and reproducible) models. Since LLM research has matured and more high-quality model families are available, have researchers adopted open-weight models? We present the first systematic study of model selection in scientific research, analyzing 21 million full-text articles through June 2026 from the Semantic Scholar Open Research Corpus (S2ORC). We employ a mixed NLP pipeline (dictionary extraction, LLM-annotation, SciBERT) to extract model occurrences in article full text and determine whether they are used or merely mentioned by researchers. We divide our corpus into single- and multi-model family studies, which we take as a proxy for applied and foundational AI research. We find that GPT-family models dominate both single- and multi-family research, but that both areas are becoming more diverse over time. In single-family papers, open-weight model use rises steadily, reaching 44.0\% in 2026. However, we find that recent growth is driven by the availability of high-quality Chinese models (e.g., Qwen, DeepSeek), which are open-weight; use of the Western open-weight families Llama and Mistral declined from 2025 to 2026. Further, a logistic regression model finds that open-weight adoption is heterogeneously distributed, estimating that researchers at Chinese institutions have 2.23 times the odds of using an open-weight model, accounting for 44.0\% of the increase in open-weight adoption since 2023, while authoring only 24.8\% of the sample in 2026. A complementary multinomial model shows that this association is concentrated in Chinese open-weight models: in 2026, their adjusted use is 37.1\% among papers with Chinese institutional affiliations, compared with 9.2\% among papers with no observed China link, a 27.9 percentage-point difference; corresponding use of other open-weight models is 15.0\% and 18.5\%, respectively. These findings suggest that open-weight adoption in science is not a general turn toward open science, but part of a broader realignment of model ecosystems in which platforms and markets, and the sociocultural and geopolitical contexts which shape them, determine which AI systems become scientific instruments.
\end{abstract}

\vspace{1.2em}
}]

\begingroup
\makeatletter
\renewcommand{\thefootnote}{}
\renewcommand{\@makefntext}[1]{\noindent #1}
\footnotetext{\raggedright
\textbf{Data and code availability.}
All materials needed to reproduce the reported results are available at
\url{https://github.com/zackarydunivin/who-uses-open-weights}.}
\makeatother
\endgroup

\section{Introduction}
Large language models (LLMs) are both objects of scientific inquiry and instruments of empirical research \citep{zheng2025automation}. Across fields, researchers study models as systems to be improved or evaluated, while also incorporating them into analytic pipelines for tasks such as classification, information extraction, synthetic data generation, behavioral experiments, and domain-specific problem solving. These uses leave traces in the published literature as experiments compare model families and applications justify the use of particular systems \citep[see, e.g.,][]{hao2026artificial,meaney2016text,goldsmithpinkham2024tracking,boeschen2023changes}. Yet despite growing evidence that LLMs are increasingly visible in scientific papers, we know little about which models researchers actually use, how those choices are changing over time, and what they reveal about the evolving infrastructure of AI and AI-enabled research.

This study is the first large-scale analysis of AI use as scientific objects rather than workflow tools, e.g., literature review, drafting, coding (but see also \citealp{hao2026artificial}). Studying researchers' model choices entails more than just counting model mentions. A model may appear in a paper as a tool, an experimental object, a related-work reference, or a generic example of contemporary AI. We present an NLP pipeline for identifying model-use claims in full-text scientific articles from S2ORC \citep{lo2020s2orc}, linked to article and author metadata from OpenAlex \citep{priem2022openalex}. The pipeline extracts candidate model mentions, classifies whether each mention reflects reported use rather than background discussion, and aggregates detected uses to model families. This allows us to track how scientific reliance on different LLM families changes over time and to ask not only which systems are most visible, but what kinds of systems researchers choose, who makes those choices, and when. Our primary analytic distinction is between proprietary and open-weight model families. Despite its coarseness, this is a consequential axis in current debates over AI-enabled science \citep{kapoor2024position,liesenfeld2024rethinking}. In the next paragraphs we explain how the open/proprietary distinction bears on scientific reproducibility, auditability, access, and control, while also emphasizing its limits.

LLM use is enabled and constrained by layered infrastructures of compute, data, interfaces, licensing, and organizational support. This complicates questions around which aspects of the apparatus can be preserved, inspected, and rerun. Reproducibility in computational research depends on access to the inputs, code, methods, and conditions under which results were produced \citep{nationalacademies2019reproducibility}. AI models are employed in science both as objects of study and tools for studying phenomena beyond AI. Open-weight availability broadens the array of possible scientific practices: trained parameters can be archived, evaluated independently, executed locally or through alternate providers, and adapted for domain-specific needs, creating affordances for scrutiny and reuse that are difficult when a model is available only as a provider-controlled service \citep{bommasani2023foundation,kapoor2024position}. For this reason, openness has become a central frame through which researchers, policymakers, and developers debate the ethics and epistemology of AI-enabled science.

At the same time, open-weight models should not be equated with open science. Recent work argues that ``open source'' claims around generative AI often mask substantial limitations, releasing model weights while withholding training data, training code, filtering procedures, documentation, access methods, or licensing freedoms \citep{liesenfeld2024rethinking,opensourceinitiative2024open}. Moreover, actual model choice is shaped by more than a release regime. Researchers encounter models through APIs, local compute constraints, linguistic and cultural communities, and commercial or state AI strategies. Policy work on AI openness accordingly treats open-weight models as part of a broader ecosystem of innovation, market structure, risk, sovereignty, and global influence \citep{oecd2025openness,meinhardt2025beyond,rand2026open}. Thus, evidence that scientists are using more open-weight models would not, by itself, indicate a simple uptake of open science practices. It also raises questions about which open-weight systems are gaining traction, where they are being used, and what infrastructural dependencies they replace or reproduce.

Our study demonstrates that scientific model choice has changed rapidly since the emergence of widely available LLMs. GPT-family models dominate early LLM-using research, but their share declines sharply as researchers incorporate a broader array of alternatives, open-weight model families among them. The rise of open-weight use, however, is not a general shift toward open models as such. In papers that use only a single model family, where model choice often reflects selection of a primary research instrument, recent open-weight adoption is concentrated in Chinese-developed model families, especially Qwen and to a lesser extent DeepSeek, while the prominent Western open-weight families, Llama and Mistral, decline slightly in multi-family papers and remain marginal in single-family papers. We further find that Chinese researchers are strongly associated with open-weight selection, even after accounting for publication date and research field. These results suggest that open-weight adoption in science should not be naively interpreted as a revealed commitment to openness or reproducibility. It also reflects the changing geography of the model ecosystem, where access, cost, performance, language coverage, institutional context, and platform availability shape which AI systems become part of scientific practice.

\section{Data and Methods}
We primarily use the Semantic Scholar Open Research Corpus (S2ORC V2) which contained 21.34 million unique structured full-text documents when pulled on July 21, 2026. While the pipeline was originally developed on data from May 7 2026, the snapshot we pulled for our production run caught S2ORC in transition to a new schema, and included duplicate and unique records across both schemas. We used all available data, deduplicating records that appeared in both, as described in further detail in the Supplementary Section~\ref{app:corpus}. 

We supplement these data with author name/affiliation and research area topic information from OpenAlex. We identified Chinese-named authors using a classifier constructed from publicly available lists of common names, which was manually augmented after large-scale testing and error audit. More detail is available in Supplementary Sections~\ref{app:corpus} \& \ref{app:metadata}.

\subsection{Determining LLM-use in Full Text}
We developed an NLP pipeline to determine which S2ORC papers report the use LLMs. A dictionary-based recognizer first extracts possible model-name occurrences from full text and maps them to model families. Separate SciBERT classifiers \citep{beltagy2019scibert} then determine whether model names actually denote the intended model and whether the paper's authors used, rather than merely mentioned, each model. Both classifiers were trained on silver-standard passages labeled by an LLM; LLM-annotation was itself validated against the authors' gold-standard annotations. A final SciBERT classifier removes extracts that resemble AI-disclosure statements, which overwhelmingly report use of LLMs as workflow tools rather than scientific instruments. Supplementary Sections~\ref{app:dictionary}--\ref{app:disclosure} provide full details.

To generate candidate occurrences of model use, we used an AI agent (Codex, GPT-5.5) to compile an initial inventory of model names, releases, developers, and access regimes. We manually verified and expanded this inventory, including by mining the ACL Anthology for omitted names. A dictionary-based extractor then searched S2ORC full text, accommodating variation in capitalization, punctuation, whitespace, version numbers, parameter sizes, and variant labels. Recognized releases were mapped to model families. 

\begin{figure*}[!t]
\centering
\includegraphics[width=0.98\textwidth]{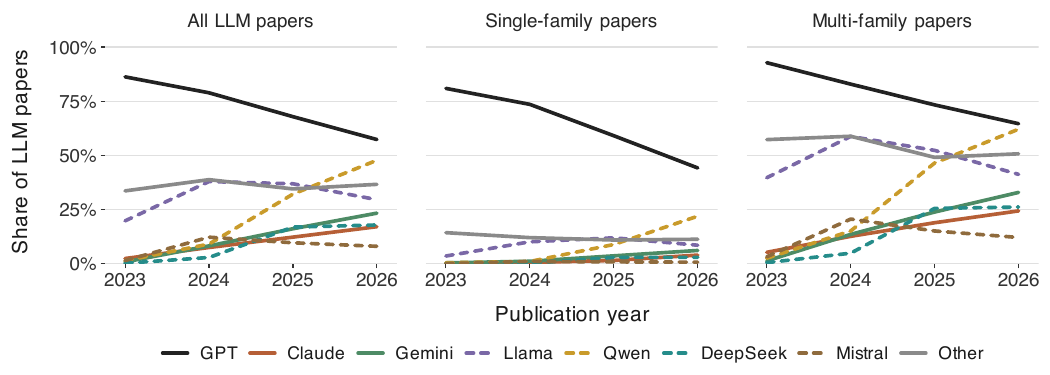}
\caption{Model-family shares among LLM-using papers, 2023--June 2026. Shares are shown for all LLM-using papers and separately for papers using a single or multiple model families. Predominately open-weight families are denoted by dashed lines.}
\label{fig:family_share}
\end{figure*}

A model name in article full text is not by itself evidence that the authors used an LLM. A matched string may not refer to a model at all. \emph{Solar}, for instance, represents solar energy in most papers, rather than denoting the language model, SOLAR. Even a genuine model reference may appear only in background discussion or related work, rather than indicating the model was used by the authors to conduct research. We therefore separate two questions: i) does the occurrence identify the intended model family, and ii) did the paper's authors use that model?

Many model names are sufficiently distinctive that a dictionary match provides strong evidence of identity. Others, such as Gemini, Falcon, Llama, and OPT, can also denote ordinary words, genes, animals, or unrelated technical systems. For a manually determined set of such ``collision-prone'' families, we required a model-identity classifier to confirm that the matched text referred to the intended LLM family.

A second classifier distinguished model use from mere mention. We classified an occurrence as use when authors reported training, fine-tuning, evaluating, or otherwise employing a model as a research instrument or object of study.

We constructed both classifiers using the same scalable strategy. As a substitute for crowdsourced training labels, we used an LLM to construct a silver-standard data set \citep{tornberg2025large, pangakis2024knowledge}. The authors annotated small gold-standard validation sets (150 passages), and then used an LLM (DeepSeek-V4-Flash) to annotate larger samples of passages, after validating against the gold standard \citep{dunivin2025scaling}. LLM-annotated silver-standard labels were then used to fine-tune SciBERT for corpus-scale classification. In both SciBERT tasks, the target occurrence was masked so that the classifier learned from contextual evidence rather than memorizing particular model names.

For model identity, DeepSeek labeled a stratified sample of 1,500 occurrences, comprising 968 genuine occurrences and 524 collisions, with 8 uncertain. This sample was drawn across collision-prone families and contextual conditions. Its labels achieved an F1 of 0.97 against the comparable gold-standard annotations. SciBERT achieved 0.92 accuracy and 0.94 F1 for identifying genuine LLM references.

For author use, DeepSeek labeled 1,500 passages, comprising 1,110 uses and 390 non-uses. Its annotations achieved 0.94 use-class F1 against 150 gold-standard passages. Five-fold cross-validation yielded 0.89 accuracy and 0.93 use-class F1 for SciBERT. We applied the trained classifiers to the extracted occurrences and retained a paper--family match when at least one occurrence satisfied the applicable identity rule and was classified as author use.

Finally, we trained a third SciBERT classifier to identify AI-disclosure statements. Such passages overwhelmingly described LLM use, largely ChatGPT, for manuscript preparation and other research-workflow tasks, including literature search and code generation. Manual review of 150 disclosure passages found that only 9\% described model use as a scientific instrument or object. We mined section headings for variants of AI-disclosure headings and treated model-mention passages under them as positive training examples; negative cases were sampled from extracts that passed the model-identity and author-use checks. In five-fold cross-validation the classifier achieved 0.98 accuracy and 0.96 disclosure-class F1. See Supplementary Section~\ref{app:disclosure} for details.

The final data set of 157,446 LLM-use papers contains 2,276,136 model-use occurrences from January 2023--June 2026 (see Supplementary Table~\ref{tab:sample_flow}).

\begin{figure*}[!t]
\centering
\includegraphics[width=0.98\textwidth]{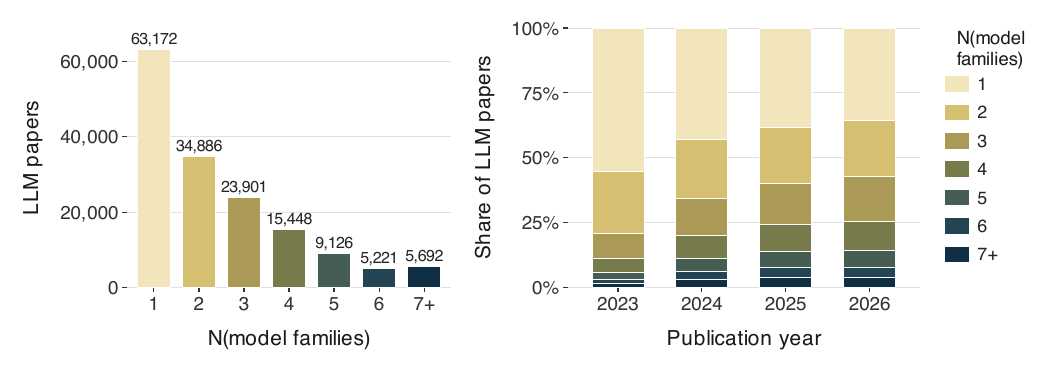}
\caption{Distribution of model-family counts among LLM-using papers. The left panel shows the number of papers by count of distinct model families used. The right panel shows the yearly share of papers in each count category.}
\label{fig:family_count_distribution}
\end{figure*}

\section{Results}
\subsection{GPT dominates early, field subsequently diversifies}
Across the corpus of papers that use LLMs, model selection diversified considerably between 2023 and 2026 (January--June). Due to its first-mover advantage \citep{lieberman1988first,arthur1989competing}, GPT-family models dominated early, appearing in a large majority of papers. Over time, however, GPT's share declined as researchers began using a wider set of model families (Figure \ref{fig:family_share}, left panel). This shift reflects the broadening array of competitive models reflecting different costs, capabilities, and values. As the model ecosystem expands, researchers face a more meaningful choice among models, including serious competition between proprietary and open-weight systems. These aggregate trends motivate a closer look at which kinds of papers and researchers are making those choices.

\subsection{Distinguishing model use from model comparison: Single- vs. multi-family papers}
The aggregate trend in Figure \ref{fig:family_share} (left panel) pools together papers that use models in different ways. Some papers employ a single model family to answer a substantive, applied, or methodological question. Others use multiple model families because comparison is itself part of the research design, as in benchmarking, evaluation, or foundation-model analysis. We therefore distinguish between \emph{single-family} papers, in which all detected model uses belong to one model family, and \emph{multi-family} papers, which use two or more distinct model families. We assume that single-family papers generally do not compare models within that family, but this is difficult to parse because it is common to refer to the same model by family name (e.g., GPT), major release (e.g., GPT-3), flavor (e.g., GPT-5 mini), or exact model version (e.g., \texttt{gpt-4-0613}). The cleavage between single- and multi-family meaningfully structures the corpus: 40.1\% of papers use a single family and 59.9\% use multiple families. Multi-family papers also become more common over time, as the single-family share falls from 55.1\% in 2023 to 35.7\% in 2026 (Figure \ref{fig:family_count_distribution}, right panel).

We treat this distinction as a crude but useful proxy for different kinds of LLM-related research. Single-family papers are more likely to capture applied or task-oriented model use, whereas multi-family papers are more likely to be foundational AI research. The context of model selection suggests different interpretations in the two settings. In multi-family papers, authors are inclined to include both open and proprietary systems because the goal entails breadth of comparison. In single-family papers, the choice is narrower. Here, researchers select one model or model family that fits their priorities and constraints. These papers therefore provide a cleaner setting for studying model selection as a revealed choice among competing model ecosystems.

\input{tables/open-weight_use_2026.tex}

\subsection{Model choice differs between single- and multi-family papers}

Returning to Figure \ref{fig:family_share}, we can see distinct adoption patterns among single- and multi-family papers. In single-family papers, proprietary GPT-family models dominated 2023, but their share declined sharply from 80.9\% in 2023 to 44.2\% in 2026 (Figure \ref{fig:family_share}, middle panel). At the same time, Qwen, a predominantly open-weight family, rose quickly, reaching 22.0\% of single-family papers in 2026. 

The trend in single-family papers reflects shifting choices among researchers, which may reveal broader drivers of model preference. In single-family papers, model choice is more tightly coupled to a single set of practical and epistemic constraints: accessibility, cost, language-competence, performance, institutional context, and values around openness. The sharp rise of Qwen in this setting indicates that open-weight adoption is not simply a diffuse turn toward all open-weight models, but may be concentrated in particular model ecosystems. By contrast, Llama, Meta's open-weight model, peaked at 12.0\% in 2025 before falling to 8.6\% in 2026, while Mistral remained marginal at 0.8\%. Among proprietary alternatives, Gemini and Claude rose to 6.2\% and 4.0\%, respectively.

Multi-family papers show a different pattern (Figure \ref{fig:family_share}, right panel). These papers are as a group more diverse throughout the period as expected because each individual paper is itself diverse. GPT remains common in this group, likely because it serves as a canonical proprietary baseline, but Qwen nearly matches it in 2026, appearing in 62.1\% of multi-family papers compared with GPT's 64.6\%. Proprietary families Gemini and Claude rise to 32.9\% and 24.4\%, respectively, while DeepSeek reaches 26.2\%. However, the uptake of open-weight models is inconsistent. Llama falls from a peak of 58.7\% in 2024 to 41.3\% in 2026, and Mistral, the only major European LLM, declines from 20.6\% to 12.1\% over the same period. The ``Other'' category, which contains many academic and earlier open-weight models, also declines after 2024, but remains present in half of multi-family papers in 2026.


\subsection{Open-weight adoption is concentrated in Chinese models}

The increase in open-weight use is substantial. Table \ref{tab:open_weight_2026} shows that in 2026, 44.0\% of single-family papers used an open-weight model family, while 87.2\% of multi-family papers included at least one such family. Among multi-family papers, open-weight models made up 59.3\% of all model-family uses. This demonstrates that open-weight models have become a major component of LLM research.

This trend does not, however, reflect a general shift toward open-weight models as a class. It is concentrated in a small set of model families. In single-family papers, Qwen appears in 22.0\% of papers, accounting for 49.9\% of all open-weight model selections. By comparison, Llama appears in 8.6\%, DeepSeek in 3.0\%, and Mistral in 0.8\%. Google's open-weight model Gemma, released in February 2024, is used in 1.9\%, while OpenAI's open-weight counterpart, GPT-OSS, released in August 2025, appears in 0.7\% of papers. 

Thus, in single-family papers, open-weight adoption is concentrated in Chinese model ecosystems, though less overwhelmingly than the earlier extraction suggested. Qwen comprises nearly half of coded open-weight selections. Along with DeepSeek and others, Chinese models account for 60.5\% of open-weight use. The rise of open-weight models in this group therefore should not be interpreted as a diffuse movement toward open science or model transparency. It is more precisely a shift disproportionately toward Chinese-developed model families, which are open-weight.



Multi-family papers show a related but distinct pattern. Open-weight families appear in 87.2\% of multi-family papers in 2026, and Qwen appears in 62.1\%. Llama and DeepSeek are also common, appearing in 41.3\% and 26.2\% of multi-family papers, respectively. However, because multi-family papers often compare many models, inclusion in this setting does not necessarily indicate that authors selected an open-weight model as their primary research instrument. Rather, these papers reflect the changing set of models that researchers consider necessary to evaluate, benchmark, or compare. Even here, however, the open-weight market is increasingly shaped by Chinese families.

\begin{figure*}[ht]
\centering
\includegraphics[width=\linewidth]{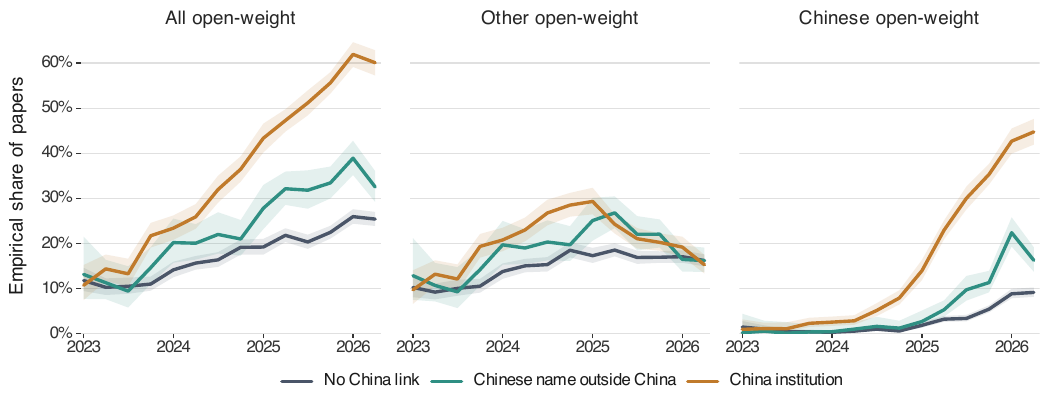}
\caption{Single-family paper open-weight model adoption by Chinese-linked authorship, 2023--June 2026.}
\label{fig:open_weight_timeseries}
\end{figure*}

\subsection{China-linked authors are more likely to select (Chinese) open-weight models}
The concentration of open-weight adoption in Qwen suggests that the rise in open-weight use is disproportionately associated with China-linked researchers. To test this, we model open-weight selection among single-family papers, where model choice is less confounded than the design of model-comparison studies. The outcome is whether the paper's sole model family is open-weight. The main predictors are Chinese institutional affiliation and Chinese-named authors, recorded for either the first or last author. In the analytic corpus of 48,129 single-family papers, 28.9\% are affiliated with a Chinese institution, 11.5\% have a Chinese-named author outside China, and 59.6\% have no China link.

Table \ref{tab:open_model_regression} reports a sequence of logistic regression models. All models include a natural spline for publication date, allowing open-weight adoption to change flexibly over time. Model 1 estimates a coefficient for Chinese institutional affiliation. Model 2 adds random intercepts for the 83 OpenAlex subfields found in the corpus, accounting for differences across research areas. Model 3 uses mutually exclusive China-link groups, distinguishing papers with Chinese institutional links from papers with Chinese-named first or last authors outside China.

Across specifications, Chinese institutional affiliation is strongly associated with open-weight selection. Model 1 estimates an odds ratio (OR) of 2.87, which declines to 2.13 after adding subfield random effects in Model 2, suggesting Chinese researchers are disproportionately likely to publish in fields that favor open weights. In Model 3, Chinese institutional affiliation has an OR of 2.23, while Chinese-name authorship is also positive, with an OR of 1.23. Model fit improves substantially after adding subfield random effects, with AUC increasing from 0.67 in Model 1 to 0.75 in Model 2. The addition of Chinese-named authors outside China shows a comparatively small improvement to model fit.

Among single-family papers in the analytic subset, which excludes papers lacking the required author, affiliation, date, or model-openness data, open-weight model adoption rose from 12.8\% in 2023 to 35.8\% in the 2026 partial-year endpoint (January--June 2026). The left panel of Figure \ref{fig:open_weight_timeseries} shows empirical adoption rates over time. Open-weight adoption rises for all groups, but it is consistently higher among Chinese-linked papers and especially high among papers with China-based institutional affiliations. A decomposition analysis (Supplementary Information, Figure~\ref{fig:chinese_link_open_adoption_decomposition}) decomposes the 23.1 percentage-point increase from 2023 to 2026 into within-group adoption and compositional changes in the sample for the two China-linked groups and the unlinked group. Within-group adoption accounts for 23.9 percentage points of the increase, compared with $-0.8$ points from composition. Papers with a Chinese institutional affiliation account for 10.2 points of the increase and Chinese-named authors outside China for 3.7 points; together, China-linked papers account for 13.9 points, compared with 9.2 points among papers without an observed China link. Thus, China-linked papers account for 60.2\% of adoption growth while constituting 38.9\% of the sample. Because remaining institutional missingness is concentrated among recent papers with Chinese-name authors, the negative composition component likely reflects differential coverage and should be interpreted cautiously (see Supplementary Section~\ref{app:metadata}).

\input{tables/open_model_regression.tex}

The binary models establish whether China-linked papers are more likely to select an open-weight family, but they treat all open-weight systems as a single category. They therefore cannot distinguish a general association with open-weight selection from adoption concentrated in model families developed in China. The middle and right panels of Figure~\ref{fig:open_weight_timeseries} show differential adoption of non-Chinese and Chinese models by each author cohort. To rigorously examine this pattern, we estimate a complementary multinomial model that classifies each paper's sole model family into three mutually exclusive categories: proprietary, other open-weight, and China-developed open-weight. As in the previous models, we make the same smooth adjustment for publication date and estimate OpenAlex subfield fixed effects. The subfield fixed effects serve the same adjustment purpose as the subfield term in the binary model while keeping the multinomial specification straightforward and estimable.

\begin{figure*}[ht]
\centering
\includegraphics[width=\linewidth]{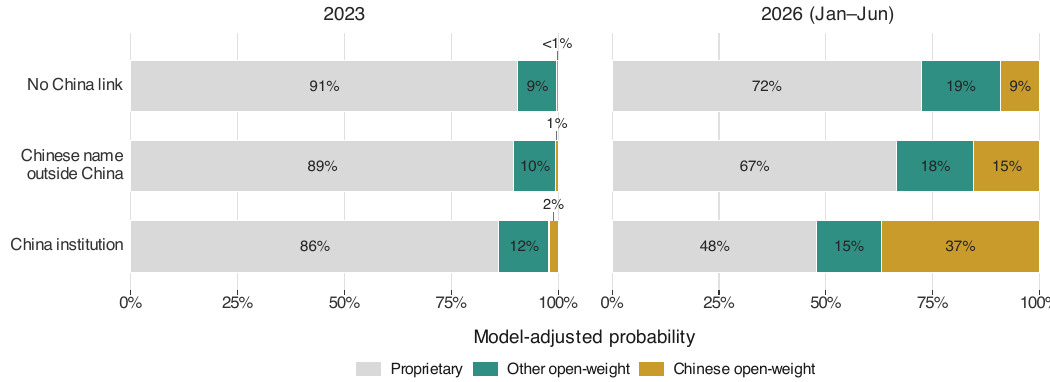}
\caption{Predicted single-family paper open-weight model adoption by Chinese-linked authorship, 2023 vs. 2026.}
\label{fig:single_family_multinomial}
\end{figure*}

Figure~\ref{fig:single_family_multinomial} reports model-adjusted probabilities for 2023 and 2026, standardized over the distribution of subfields. The model estimates a association between China linkage and model choice over the study period. Separating out the year-specific probabilities illustrates how that association translates into absolute differences as the availability and use of Chinese-developed models change over time.

The temporal comparison shows that the association of open-weight families with Chinese institutions is concentrated in Chinese models rather than open-weight models generally. In 2023, the adjusted probability of selecting a Chinese open-weight family was low in every group: 0.3\% among papers with no observed China link, 0.6\% among papers with Chinese-named authors outside China, and 2.2\% among papers with a Chinese institutional affiliation. By 2026, these probabilities had risen to 9.2\%, 15.4\%, and 37.1\%, respectively. Thus, at the June 2026 endpoint, papers with a Chinese institutional affiliation were 27.9 percentage points more likely than papers with no observed China link to select a Chinese open-weight family. The corresponding probabilities for other open-weight families were 15.0\% and 18.5\%, respectively. The weaker name-based association outside China was similarly concentrated in Chinese open-weight families. These papers were 6.2 points more likely to select a Chinese open-weight family in 2026, while their probability of selecting another open-weight family did not differ from that of papers with no observed China link. Thus, the broad institutional association with open-weight selection in the logistic model is driven primarily by adoption of Chinese model families.

\section{Discussion}
Our results show that the recent rise of open-weight model use in science is not a generalized turn toward open science. In 2026, 61\% of open-weight selections in single-family papers were Chinese-developed models, 83\% of which is attributable to Qwen. Our institutional analysis sharpens this finding. Papers affiliated with Chinese institutions were substantially more likely to select Chinese-developed open-weight families, but not other open-weight families. If openness itself drove this institutional pattern, China-affiliated papers should also select non-Chinese open families at higher rates; they do not. Aggregate open-weight adoption therefore records the rise of particular model ecosystems, not the diffusion of openness as a scientific norm.

Researchers do not select openness or closure in the abstract. They choose among systems that bundle capability, price, language performance, and institutional factors. The contrasting trajectories of Qwen and DeepSeek on one hand and Llama and Mistral on the other suggest that releasing weights translates into scientific use only when attached to a competitive technical and economic package. Open release is itself an industrial strategy. It can attract developers, stimulate downstream complements, establish standards, and weaken incumbent API platforms \citep{oecd2025openness,meinhardt2025beyond,rand2026open}. Open-weight growth may therefore reflect competition among model producers more than a change in scientists' methodological values or a realignment of choices with pre-existing values.

LLMs are increasingly becoming infrastructure for scientific work. Chinese open-weight models offer alternatives to a small set of U.S. proprietary providers and broaden access to capable models. As researchers move from GPT toward Qwen or DeepSeek, however, scientific dependence is being distributed rather than dissolved. But what distribution we see is fairly limited. Our results show both global diffusion and institutional embeddedness. Chinese models are increasingly used outside China, yet adoption remains much stronger within Chinese institutional settings. Model artifacts can circulate globally while uptake remains structured by geography, language, professional networks, and domestic technology ecosystems. Individually pragmatic and institutionally situated choices can thus aggregate into a geopolitical realignment of scientific infrastructure even if openness itself plays little role in why particular models are selected.

These findings also caution against romanticizing open weights as open science. Downloadable parameters enable local execution, preservation, adaptation, and reduced exposure to unilateral API changes, but they do not make models transparent in the ordinary scientific sense. Training data and development procedures often remain unavailable, model behavior remains extremely difficult to explain, and technical reproducibility neither creates incentives for replication nor ensures that it occurs \citep{liesenfeld2024rethinking,opensourceinitiative2024open}. We cannot identify researchers' motives or separate the effects of performance, price, language, access, and institutional support. Corpus coverage and authorship proxies also limit precision in this analysis. The industrial field is developing quickly, and published papers will show a lag in new tool use: GPT-OSS and Kimi 2.6+ models should be expected to rise in the coming year. Even so, the lesson of this study is that the relevant unit for understanding AI-enabled science is not an abstract open--proprietary binary, but the industrial and geopolitical ecosystem in which a model is produced, distributed, and used. The future of openness in AI and AI-enabled science depends less on scientists' commitment to openness than on which model ecosystems make openness competitive.

\section*{Acknowledgments}
We thank Patrick Kaminski, Raphael Heiberger, and Harry Yan for comments contributing to the improvement of this manuscript. 
\bibliographystyle{zod_sorted_nat}
\bibliography{references}


\clearpage
\appendix
\setcounter{section}{0}
\setcounter{subsection}{0}
\setcounter{table}{0}
\setcounter{figure}{0}
\setcounter{equation}{0}
\renewcommand{\thesection}{S}
\renewcommand{\thetable}{S\arabic{table}}
\renewcommand{\thefigure}{S\arabic{figure}}
\renewcommand{\theequation}{S\arabic{equation}}
\section*{Supplementary Information}
\input{technical_appendix}

\end{document}

%% file: tables/open-weight_use_2026.tex
\begin{table*}[!htb]
\centering
\caption{Open-weight model-family use in 2026 LLM-using papers. Percentages in the paper columns are shares of papers in each family-count group. The denominator is 16,125 single-family papers and 29,000 multi-family papers. The multi-family use percentage takes the number of model families used in each paper as the denominator ($N=102,476$). The final row is a union over the listed open-weight families.}
\label{tab:open_weight_2026}
\begin{tabular}{lrrrrr}
\toprule
& \multicolumn{2}{c}{Single-family papers} & \multicolumn{3}{c}{Multi-family papers} \\
\cmidrule(lr){2-3}\cmidrule(lr){4-6}
Model family & $n$ & \% papers & $n$ & \% papers & \% family uses \\
\midrule
Qwen & 3,544 & 22.0 & 18,011 & 62.1 & 17.6 \\
Llama & 1,382 & 8.6 & 11,963 & 41.3 & 11.7 \\
DeepSeek & 482 & 3.0 & 7,599 & 26.2 & 7.4 \\
Gemma & 311 & 1.9 & 4,248 & 14.6 & 4.1 \\
Mistral & 123 & 0.8 & 3,520 & 12.1 & 3.4 \\
GPT-OSS & 105 & 0.7 & 1,868 & 6.4 & 1.8 \\
Kimi & 19 & 0.1 & 1,382 & 4.8 & 1.3 \\
Other open-weight & 1,131 & 7.0 & 9,059 & 31.2 & 11.9 \\
\midrule
\textbf{Any coded open-weight family} & \textbf{7,097} & \textbf{44.0} & \textbf{25,291} & \textbf{87.2} & \textbf{59.3} \\
\bottomrule
\end{tabular}
\end{table*}

%% file: tables/open_model_regression.tex
\begin{table}[t]
\caption{Logistic regression of China-linked authorship on open-model selection in single-family papers. All models include a natural spline for publication date. Institution and name measures are restricted to first and last
authors. Models 2 and 3 include random intercepts for OpenAlex subfield. Coeffients are given as odds ratios.}
\label{tab:open_model_regression}
\centering
\begingroup
\def\arraystretch{1.25}
\newcommand{\sym}[1]{\ifmmode^{#1}\else\(^{#1}\)\fi}
\sisetup{
    detect-all,
    group-digits             = true,
    group-minimum-digits     = 4,
    group-separator          = {,},
    table-align-text-post    = false,
    table-space-text-post    = {$^{***}$}
}
\begin{tabular}{p{0.32\linewidth}ccc}
\toprule
 & \multicolumn{1}{c}{Model 1}
 & \multicolumn{1}{c}{Model 2}
 & \multicolumn{1}{c}{Model 3} \\
\midrule
China Institution
    & 2.87\sym{***} 
    & 2.13\sym{***} 
    & 2.23\sym{***} 
    \\
Chinese Name Outside China
    & \multicolumn{1}{c}{---}
    & \multicolumn{1}{c}{---}
    & 1.23\sym{***} 
    \\
\midrule
Subfield RI levels
    & \multicolumn{1}{c}{---}
    & \multicolumn{1}{c}{83}
    & \multicolumn{1}{c}{83} \\
Observations
    & \multicolumn{1}{c}{48,129}
    & \multicolumn{1}{c}{48,129}
    & \multicolumn{1}{c}{48,129} \\
AIC & 51710 & 48169 & 48137 \\
AUC & 0.674 & 0.748 & 0.749 \\
\bottomrule
\end{tabular}
\begin{minipage}{0.95\linewidth}
\vspace{0.4em}
\footnotesize\sym{*} \(p<.05\); \sym{**} \(p<.01\); \sym{***} \(p<.001\).
\end{minipage}
\vspace{-0.4em}
\endgroup
\end{table}

%% file: technical_appendix.tex

\begin{table*}[t]
\centering
\small
\caption{Construction of the analytic samples. Occurrences are non-reference
model-name matches in canonical S2ORC records. Paper--family pairs collapse
repeated qualifying occurrences of the same family within a paper. The first
row counts source records; subsequent rows count unique papers after
deduplication by \texttt{CorpusId}.}
\label{tab:sample_flow}
\begin{tabular}{lrrr}
\toprule
Stage & Papers & Occurrences & Paper--family pairs \\
\midrule
S2ORC source records
    & 35,599,352 & --- & --- \\
Unique record CorpusIds
    & 21,338,177 & --- & --- \\

\addlinespace
\multicolumn{4}{l}{\emph{Model-use classification}} \\
Model-name occurrence
    & 339,787 & 3,619,386 & 683,929 \\
Pass model-identity rule
    & 220,844 & 3,213,035 & 560,822 \\
Pass previous and author-use rules
    & 178,531 & 2,366,252 & 417,293 \\
Pass previous and AI-disclosure rules
    & 167,451 & 2,346,686 & 402,736 \\

\addlinespace
\multicolumn{4}{l}{\emph{Dated analytic corpus}} \\
Usable publication year
    & 167,236 & 2,343,502 & 402,227 \\
Published January 2023--June 2026
    & 157,446 & 2,276,136 & 389,070 \\
\quad Single-family
    & 63,172 & 379,177 & 63,172 \\
\quad Multi-family
    & 94,274 & 1,896,959 & 325,898 \\

\addlinespace
\multicolumn{4}{l}{\emph{Institutional analysis: single-family papers}} \\
Known access regime
    & 62,448 & 377,311 & 62,448 \\
\quad OpenAlex author records
    & 56,854 & 354,693 & 56,854 \\
\quad First- or last-author country
    & 48,129 & 315,155 & 48,129 \\
Final regression sample
    & 48,129 & 315,155 & 48,129 \\
\quad Dated January 1
    & 4,001 & 27,135 & 4,001 \\
\bottomrule
\end{tabular}
\end{table*}

\subsection{Corpus Construction}
\label{app:corpus}




\subsubsection{Sample construction}

The S2ORC snapshot captured the corpus during a transition between two record formats. These two versions largely overlapped in coverage, but a substantial fraction of papers appeared in only one.

We therefore treated the Semantic Scholar \texttt{CorpusId} as the stable paper identifier. When a paper appeared in both formats, we retained the richer body-schema record and ignored the older copy. When it appeared in only one format, we retained the available record. We did not combine model matches across versions of the same paper. Of 35,599,352 source records, 14,261,151 \texttt{CorpusId}s appeared in both components. A further 24 records were repeated within the richer component. Deduplication produced 21,338,177 unique papers: 2,989,168 found only in the older format, 4,087,858 found only in the richer format, and 14,261,151 represented in both.

\paragraph{Record linkage and exclusions.} S2ORC records were linked to OpenAlex by DOI. DOI strings were normalized by removing URL and \texttt{doi:} prefixes and converting to lowercase before lookup. The enrichment pipeline first retained previously linked OpenAlex identifiers and then queried OpenAlex by S2ORC DOI when an identifier was absent. It matched 117,324 of 127,643 papers (91.9\%). OpenAlex returned at least one author record for 117,281 matched papers, but only 30,115 had a structured institutional affiliation and 30,091 had a country-coded institution. The binary regression then retained papers published from 2023 through the corpus cutoff that used exactly one model family and had country information for at least one of the first or last authors. These restrictions yielded 48,129 papers. Missing subfields were retained as an ``Unknown'' category.



\subsection{Author, Affiliation, and Field Measures}
\label{app:metadata}

\subsubsection{Name-origin classifier}

We could not find a publicly available classifier for Romanized Chinese names. Instead, we constructed our own name-origin classifier through collaboration with an AI-agent (Codex, GPT 5.5). The AI agent compiled candidate surnames, romanization variants, and Pinyin syllables from public lists and implemented the initial rules. The author reviewed recurrent false positives and false negatives surfaced by corpus-scale diagnostics, decided which additions, exclusions, and ambiguity guards were substantively justified, and instructed the AI agent to encode and retest those revisions. The resulting rule set is used only as a noisy indicator of a possible connection to a linguistic or cultural ecosystem when no China-based institutional affiliation is observed. It does not identify nationality, citizenship, ethnicity, self-identification, or location.

Names shared across regions or represented through different transliteration conventions can be misclassified. We therefore report institutional affiliation and name-origin associations separately and interpret the latter cautiously. The rules combine public common-Chinese-surname tables, compiled Chinese-surname lists, a standard Pinyin-syllable table, and additional Cantonese/Hong Kong, Hokkien/Teochew, Hakka, Wade--Giles, Southeast Asian, and compound-surname variants. Text is Unicode-normalized, lowercased, and stripped of diacritics before tokenization. Ambiguous surnames require supporting given-name evidence; a small audited non-Chinese given-name blocklist and Korean-surname suppression reduce recurrent false positives. In a diagnostic comparison with 500,000 high-confidence \emph{weak-labeled} names sampled from OpenAlex, the strict rule achieved 0.898 precision, 0.911 recall, 0.977 specificity, and 0.965 accuracy relative to those labels. Because the reference labels were themselves heuristic rather than human-adjudicated identities, these statistics measure agreement with the diagnostic labels and should not be interpreted as population-valid classification accuracy. The classifier is available in the associated materials.

\begin{table*}[ht]
\centering
\small
\caption{First- or last-author institutional-country coverage before and after OpenAlex author profile recovery. The denominator is the 62,448 known access-regime single-family papers published from 2023 through June 2026.}
\label{tab:institution_coverage_recovery}
\begin{tabular}{lrrr}
\toprule
Paper group & Work authorship & Exact publication year &
Exact year + last-known \\
\midrule
All papers & 42.6\% & 77.1\% & 78.2\% \\
Open-weight family & 25.9\% & 65.9\% & 67.0\% \\
Proprietary family & 50.1\% & 82.1\% & 83.2\% \\
Chinese-name endpoint author & 34.6\% & 83.5\% & 84.2\% \\
Other classified endpoint names & 54.9\% & 85.4\% & 87.0\% \\
Endpoint-name classification missing & 0.0\% & 0.0\% & 0.0\% \\
Qwen & 12.4\% & 49.6\% & 50.1\% \\
2026 Qwen & 8.9\% & 39.5\% & 39.5\% \\
\bottomrule
\end{tabular}
\end{table*}

\subsubsection{China-link categories}

The regression analysis partitions papers into three mutually exclusive categories using the first and last authors:
\begin{enumerate}
\item \textbf{Chinese institutional affiliation}: at least one endpoint author has an affiliation located in China;
\item \textbf{Chinese name outside China}: no observed Chinese institutional affiliation, but at least one endpoint-author name is classified as Chinese-origin; and
\item \textbf{No China link}: neither condition is observed.
\end{enumerate}
In the 48,129-paper binary-regression sample, these groups comprise 28.9\%, 11.5\%, and 59.6\% of papers, respectively.

First and last authors were determined from OpenAlex authorship positions. A paper had a Chinese institutional link when \texttt{CN} appeared among the country codes of either endpoint author's institutions. Papers were eligible for the institutional analyses when country data were available for at least one author. When only one had country data, we used the available information.

\subsubsection{Affiliation sources and author-profile recovery}

Where available, we used institutions attached to each authorship on the focal OpenAlex work. This source has the strongest temporal connection to the paper, but country data were frequently absent. We therefore to recover missing first- and last-author countries from OpenAlex author profiles.

The main specification uses only profile affiliations whose \texttt{years} field contains the paper's publication year. If multiple qualifying affiliations had country codes, we retained all observed countries. OpenAlex's \texttt{last-known institution} field was used when publication-year affiliation was absent, but only in a sensitivity analysis in this appendix. 

We assessed 53,785 missing endpoint-author slots, corresponding to 43,298 unique OpenAlex author profiles when an author identifier was available. OpenAlex returned 43,286 profiles. Exact-year profile affiliations recovered 34,982 endpoint-author slots across 22,182 papers. Last-known institutions recovered 2,648 additional slots.

\subsubsection{Residual coverage and selection}

Author-profile recovery substantially reduced selection on affiliation recorded in the OpenAlex article records. The open-weight--proprietary coverage gap fell from 24.2 to 16.2 percentage points, and the pooled gap between Chinese-name and other name-classified papers fell from 20.3 to 1.9 points. These pooled comparisons, however, conceal substantial residual selection in the newest papers.

Because name classification does not require institutional country data, it provides an internal diagnostic for affiliation missingness among papers with classifiable endpoint names. Through 2025, exact-year country coverage was similar or higher among papers with a Chinese-name endpoint author. In 2026, however, coverage was only 62.8\% for Chinese-name papers, compared with 77.2\% for papers with other classified names (Table~\ref{tab:institution_coverage_by_name_year}).

The 2026 deficit is not equally distributed across institution and model accessibility regimes. Within 2026, coverage was 56.0\% for open-weight papers with a Chinese-name endpoint author and 72.4\% for proprietary papers with a Chinese-name endpoint author; the corresponding rates for other classified names were 67.2\% and 81.4\%. Coverage was therefore lower both for open-weight papers and for papers with Chinese-name endpoint authors. Qwen papers are an especially visible intersection of these patterns, but the name-based diagnostic shows that the problem is not confined to one model family.

The data do not identify why recent Chinese-name authors are less likely to have recoverable institutional countries. Possible mechanisms include delayed indexing of recent publications, incomplete work-to-author linkage, and sparse or temporally incomplete author profiles. It is likely that this reflects a higher rate of entry for Chinese researchers, though it may also stem from OpenAlex's sourcing practices.

All institutional models are consequently conditional on observed endpoint-author country and name information. Residual missingness is associated with publication year, name classification, and model openness, so it is not missing completely at random. The decomposition's sample-composition component is particularly exposed because it depends directly on the observed shares of the three China-link groups. In the main decomposition, the China-institution group's within-group adoption component is $+12.1$ percentage points, while its changing observed sample share contributes $-1.9$ points, for a net contribution of $+10.2$ points. The negative composition component should not be interpreted as evidence that the population share of China-affiliated papers declined: it may partly reflect lower institutional coverage in the recent cohort.

\begin{table}[t]
\centering
\small
\caption{Exact-year first- or last-author country coverage by publication year and endpoint-name classification.}
\label{tab:institution_coverage_by_name_year}
\begin{tabular}{lrr}
\toprule
Year & Chinese-name & Other classified names \\
\midrule
2023 & 92.9\% & 91.1\% \\
2024 & 92.1\% & 88.4\% \\
2025 & 88.7\% & 86.1\% \\
2026 & 62.8\% & 77.2\% \\
\bottomrule
\end{tabular}
\end{table}

\subsubsection{Last-known-institution sensitivity}

Adding last-known institutions increased endpoint-country coverage by only one percentage point, from 77.1\% to 78.2\%. The substantive estimates were nearly unchanged. In the final binary model, the Chinese-institution odds ratio was 2.23 [2.12, 2.34] under exact-year recovery and 2.19 [2.09, 2.30] when last-known institutions were also admitted. The corresponding Chinese-name odds ratios were 1.23 [1.15, 1.32] and 1.24 [1.16, 1.33]. In the multinomial model, the Chinese-institution contrasts were 13.7 versus 13.5 percentage points for Chinese open-weight families and 0.5 versus 0.4 points for other open-weight families. The China-institution decomposition contribution was 10.2 points in both specifications.

This stability is reassuring about the narrow choice between exact-year and last-known profile evidence. It does not eliminate the broader missing-data concern. The sensitivity adds little coverage in the problematic recent cohort, including virtually no improvement for 2026 Qwen papers. We therefore use exact-year affiliations as the main measure, report last-known institutions only as a sensitivity analysis, and interpret results involving changes in observed group composition cautiously.

\subsubsection{Research subfields}

We use the subfield of each work's OpenAlex primary topic, falling back to the previously joined metadata field when necessary. We observed 230 subfields. In the binary regression (see Appendix~\ref{app:models}), subfields represented by fewer than 50 papers were pooled as ``Other rare subfield,'' and missing values were labeled ``Unknown,'' producing 83 random-effect levels; rare subfields were not pooled in the multinomial model. Subfield adjustment accounts for the likelihood that Chinese researchers and open-weight use are differently distributed across research areas.

Table~\ref{tab:openalex_topic_fields_percentages} lists the distribution of each OpenAlex field for both single- and multi-family papers. Computer Science is the most frequent field in both groups, but more common among multi-family papers (70.9\% primary topic) than single-family papers (55.9\%). This is consistent with the assumption that single-family papers are more likely to represent applied rather than foundational AI research. Across both paper groups the other leading subfields are Medicine, Social Sciences (which excludes Psychology and Economics), Engineering, and Psychology. Although we model subfield, not field, we display field because it is easier to interpret and roughly as informative as the subfield breakdown. Table~\ref{tab:openalex_topic_fields_percentages} also gives the number of observed subfields per field. 

\begin{table*}[t]
\centering
\small
\setlength{\tabcolsep}{8pt}
\caption{OpenAlex field representation and number of observed subfields, by model-family count. Primary and any-topic percentages use papers with OpenAlex topic arrays as the denominator (single-family: 57\,524 of 63\,172, 91.1\%; multi-family: 83\,027 of 94\,274, 88.1\%).}
\label{tab:openalex_topic_fields_percentages}
\begin{tabular}{p{0.34\textwidth}rrrrr}
\toprule
& & \multicolumn{2}{c}{Single-family papers} & \multicolumn{2}{c}{Multi-family papers} \\
\cmidrule(lr){3-4}\cmidrule(lr){5-6}
OpenAlex field & \shortstack{No. of\\subfields} & \shortstack{Primary\\topic (\%)} & \shortstack{Any\\topic (\%)} & \shortstack{Primary\\topic (\%)} & \shortstack{Any\\topic (\%)} \\
\midrule
Computer Science & 11 & 55.9 & 74.2 & 70.9 & 85.7 \\
\rowcolor[gray]{0.92} Medicine & 41 & 12.3 & 18.5 & 7.1 & 12.6 \\
Social Sciences & 22 & 7.7 & 16.9 & 5.7 & 13.1 \\
\rowcolor[gray]{0.92} Engineering & 16 & 6.2 & 12.7 & 3.8 & 8.3 \\
Psychology & 7 & 3.9 & 9.9 & 2.9 & 7.0 \\
\rowcolor[gray]{0.92} Decision Sciences & 4 & 2.5 & 6.2 & 2.6 & 6.2 \\
Biochemistry, Genetics and Molecular Biology & 14 & 2.3 & 5.0 & 1.4 & 3.7 \\
\rowcolor[gray]{0.92} Neuroscience & 8 & 1.6 & 3.6 & 1.0 & 2.4 \\
Business, Management and Accounting & 8 & 1.4 & 3.2 & 1.1 & 2.4 \\
\rowcolor[gray]{0.92} Health Professions & 11 & 1.1 & 3.3 & 0.6 & 2.0 \\
Arts and Humanities & 13 & 1.0 & 2.6 & 0.6 & 1.8 \\
\rowcolor[gray]{0.92} Environmental Science & 11 & 0.8 & 1.6 & 0.2 & 0.5 \\
Materials Science & 7 & 0.7 & 1.7 & 0.8 & 2.2 \\
\rowcolor[gray]{0.92} Physics and Astronomy & 8 & 0.7 & 1.5 & 0.3 & 0.8 \\
Agricultural and Biological Sciences & 11 & 0.5 & 0.8 & 0.2 & 0.2 \\
\rowcolor[gray]{0.92} Economics, Econometrics and Finance & 3 & 0.5 & 1.2 & 0.3 & 0.7 \\
Mathematics & 10 & 0.3 & 0.8 & 0.2 & 0.5 \\
\rowcolor[gray]{0.92} Earth and Planetary Sciences & 8 & 0.3 & 0.8 & 0.1 & 0.3 \\
Immunology and Microbiology & 5 & 0.1 & 0.3 & 0.0 & 0.1 \\
\rowcolor[gray]{0.92} Dentistry & 4 & 0.1 & 0.2 & 0.1 & 0.2 \\
Chemistry & 6 & 0.1 & 0.3 & 0.0 & 0.1 \\
\rowcolor[gray]{0.92} Pharmacology, Toxicology and Pharmaceutics & 3 & 0.1 & 0.2 & 0.1 & 0.1 \\
Nursing & 4 & 0.1 & 0.1 & 0.0 & 0.1 \\
\rowcolor[gray]{0.92} Energy & 3 & 0.0 & 0.2 & 0.0 & 0.1 \\
Veterinary & 2 & 0.0 & 0.1 & 0.0 & 0.0 \\
\rowcolor[gray]{0.92} Chemical Engineering & 6 & 0.0 & 0.1 & 0.0 & 0.0 \\
\bottomrule
\end{tabular}
\end{table*}

\subsection{Model Dictionary, Family Mapping, and Model Metadata}
\label{app:dictionary}

\subsubsection{Model dictionary construction}

We employed an AI agent (Codex, GPT-5.5) to generate an initial inventory of model names, families, developers, and access regimes. To identify omissions, we then mined full text from the ACL Anthology (ACL, EACL, NAACL, TACL, EMNLP) for model names. ACL papers provided a useful discovery corpus because they contain many discussions of language models and their XML text is generally cleaner than the S2ORC text used in the main analysis. Candidate strings were surfaced from model-related passages and compared with the existing inventory.

The authors reviewed the resulting candidates in context, using concordances to distinguish model names from datasets, methods, software, and ordinary words. They directed additions, removals, and corrections, including the addition of missing families and common alternative names. The ACL corpus was used to construct and audit the inventory, not as the corpus from which the paper's substantive results were estimated.

Our aim was broad coverage of model families. The final inventory is therefore an author-adjudicated research instrument produced with agent assistance. Final decisions about inclusion and family assignment were made by the authors, and the machine-readable inventory accompanies the code archive. We ended up with 159 model roots that we classified as generative large language models, and organized these roots into 98 model families. The full dictionary also includes other deep-learning-based language models, e.g., embedding and BERT models, as well as model platforms and providers.

\subsubsection{Family taxonomy}
We distinguish three levels of identity. A \emph{model} is the text observed in a paper, such as \texttt{Claude 3.7 Opus}, \texttt{gpt-4o-mini}, or \texttt{Mistral-7B-Instruct-v2}. A \emph{family} is the stable analytic unit to which model are mapped, such as GPT or Mistral. A \emph{superfamily} is an optional downstream grouping of related families, such as families in the Mistral or Llama ecosystem. In the main paper we use \emph{family} to stand for both families and superfamilies.

This organization allows the extractor to recognize releases that are not individually enumerated. It first recognizes a family root and then parses adjacent version numbers, parameter sizes, and variant labels. Thus, an otherwise unknown can still be assigned to the appropriate family. Parsed release and size information is retained as metadata when available, but the principal analysis counts distinct families. Repeated occurrences and multiple releases from one family therefore contribute one member to a paper's family set. Families and super families were organized by the AI agent and manually inspected by the authors.

\subsubsection{Access-regime and developer-country coding}

We use \emph{open-weight} to describe a release whose trained parameters are available for download or local execution. This designation does not imply that its training data, training code, filtering procedures, or license meet a comprehensive open-source standard. A proprietary release is one for which the trained parameters are not publicly released and access is primarily provider-controlled. Coding reflects publicly documented availability at the analytic publication cutoff reported in Section~\ref{app:corpus}.

Access regime was coded downstream from name recognition. This separation prevents assumptions about openness from determining whether a model mention is extracted. Where the observed surface identifies a release, release-specific coding takes precedence. Family defaults are used only for families with a sufficiently consistent access regime. Families containing both downloadable and provider-controlled releases are marked mixed until a release-specific rule or an explicitly documented paper-level resolution applies. Generic mixed-family mentions are not automatically treated as open-weight. 

A China-developed open-weight family is an open-weight family whose original developer is headquartered in China.

Access regime and developer country were determined by an AI agent (Codex, GPT 5.5) for each discovered model surface and verified by the authors. The complete table, including aliases and sources for each coding decision, is available in machine-readable form.

\subsection{Model-Occurrence Extraction}
\label{app:extraction}

\subsubsection{Extraction objective}

Having defined the model dictionary, we next searched S2ORC full text for possible occurrences of its model families. The purpose of this stage was to generate candidates for classification, not to decide whether every matched string actually denoted a language model or whether the article's authors used it. Those questions require understanding the surrounding passage and are addressed by the classifiers described in Section~\ref{app:classification}.

The unit of extraction was therefore a \emph{candidate occurrence}: a particular span of article text provisionally assigned to a family in the model dictionary. This distinction allowed the extractor to favor coverage. An irrelevant occurrence of an ambiguous name could be removed later, whereas a model occurrence never extracted could not be recovered by either classifier.

\subsubsection{Text preparation and sentence segmentation}

We processed the structured sections of each S2ORC article while retaining their section labels, titles, and positions in the document. S2ORC text contains irregular line breaks, nonstandard Unicode punctuation, zero-width characters, and inconsistent spacing. We Unicode-normalized the text, removed zero-width characters, and collapsed runs of whitespace before matching.

We then divided each section into sentences and searched each sentence separately. Sentence boundaries are necessarily approximate in machine-extracted scientific text, especially in tables, references, mathematical expressions, and malformed passages. Nevertheless, sentence-level extraction provided a useful and generally interpretable unit of context for classification. It also allowed us to preserve the paper, section, and sentence location of every occurrence.

\subsubsection{Family-root recognition and surface parsing}

The extractor translated the dictionary's 159 searchable roots into flexible matching rules. Boundary guards prevented a root from matching inside a longer alphanumeric string. The matcher admitted spaces, nonbreaking spaces, underscores, slashes, hyphens, and common Unicode dash characters as separators. Most roots were matched without regard to capitalization, although configured forms such as \texttt{GPT}, \texttt{PaLM}, \texttt{LaMDA}, and \texttt{OPT} retained capitalization requirements where case helped distinguish the model name from another expression.

After identifying a family root, the extractor parsed adjacent version numbers, parameter sizes, and common variant labels. It could therefore recognize surfaces such as \texttt{GPT4}, \texttt{GPT 4.1}, \texttt{GPT-4V}, and \texttt{Llama-3.1-8B-Instruct} without requiring every complete string to be listed in advance. The parser admitted both attached and separated major or decimal releases, parameter counts, and labels such as \texttt{Chat}, \texttt{Coder}, \texttt{Instruct}, \texttt{Vision}, and \texttt{VL}.

This family-first design permits limited generalization to releases that were not individually enumerated. For example, a previously unlisted surface such as \texttt{Mistral-48B-Instruct-v2} can still be assigned to the Mistral family because its family root and suffix structure are recognizable. The parsed release, size, and variant information was retained as metadata, but the family assignment was the principal output used downstream.


\subsubsection{Candidate-entry rules for ambiguous roots}

Not every dictionary root was extracted whenever its letters appeared. Some family names are distinctive enough to be plausible candidates as bare strings, whereas others are common words, abbreviations, names, or scientific terms. We therefore assigned each family a minimum candidate-entry policy.

Of the 98 family entries, 46 admitted a configured bare root. 17 ordinarily required a suffix or an explicitly permitted alias, 27 required nearby general model terminology, and eight required more specific language-model terminology. Individual entries could also specify capitalization requirements, provider-qualified aliases, or exceptions for distinctive forms.

These rules were intended to avoid matches that were plainly implausible while preserving uncertain cases for classification. For example, \texttt{DeepSeek} could be admitted as a distinctive bare root. Bare \texttt{GPT} was not sufficient, but numbered releases and configured forms such as \texttt{InstructGPT} were candidates. \texttt{Phi} ordinarily required a numbered release or a form such as \texttt{Microsoft Phi}. Provider-qualified expressions such as \texttt{Google Bard}, \texttt{Anthropic Claude}, and \texttt{Upstage SOLAR} supplied stronger evidence than their unqualified roots.

For some ambiguous bare names, nearby words referring to models, prompting, inference, benchmarking, evaluation, or language modeling were sufficient to admit the occurrence to the candidate pool. This was deliberately a weak standard. A word such as ``model'' might make an occurrence worth examining, but it did not establish that the occurrence denoted the intended LLM. For example, a surface such as \texttt{SOLAR-10.7B-Instruct} could enter the candidate pool even though \emph{solar} has many unrelated scientific meanings. The model-identity classifier described in Section~\ref{app:classification}, rather than the extraction rule, made the later semantic decision for such collision-prone families.

\subsubsection{Contextual keyword flags}

In addition to applying the family-specific entry rules, we recorded several sentence-level context flags for every extracted occurrence. The flags identified four broad types of language:

\begin{enumerate}
    \item \emph{explicit language-model terminology}, including ``language model,'' ``large language model,'' ``LLM,'' ``foundation model,'' ``vision-language model,'' ``multimodal language model,'' ``embedding model,'' and ``generative AI'';
    \item \emph{model-operation terminology}, including fine-tuning, instruction-tuning, zero-shot or few-shot inference, chain-of-thought, tokenization, and API calls;
    \item \emph{NLP-task terminology}, including natural language processing, text generation, machine translation, question answering, summarization, dialogue systems, and chatbots; and
    \item \emph{transformer terminology}, which was recorded separately because it is relevant but much less specific to LLMs.
\end{enumerate}

We also aggregated these signals across the paper. This allowed us to distinguish occurrences accompanied by explicit LLM language in the same sentence from those supported only by language elsewhere in the article. Other occurrences appeared with operational or NLP-task language, with transformer terminology alone, or with none of the recorded contextual terms.

These keyword flags were heuristic descriptions of an occurrence's context. We later used the flags to construct a heterogeneous sample for model-identity annotation (see Section~\ref{app:classification}). The sample was stratified across contextual conditions, surface forms, collision-prone families, and whether another model name appeared in the passage. This ensured that the annotation and training data included both easy cases with explicit language-model terminology and difficult cases with little or misleading contextual evidence.

During pipeline development, we also used these signals in a preliminary paper-level evidence heuristic and wrote passing and failing candidates to separate outputs. Audits showed why that heuristic was inadequate as a final decision rule: it admitted irrelevant occurrences when an otherwise unrelated paper happened to contain model-related language, while excluding some genuine model occurrences lacking the selected terminology. We therefore recombined both sets of candidates for classification. The preliminary paper-level heuristic does not determine inclusion in the final analytic sample.

\subsubsection{Audit and limitations}

We audited extraction using concordances from ACL Anthology full text, comparisons with the earlier alias inventory, generated tests spanning capitalization and separator variants, and negative examples drawn from recurrent scientific collisions. Regression tests covered attached and decimal releases, parameter sizes, malformed whitespace, overlapping roots, capitalization-sensitive forms, and ambiguous strings used outside language-model research. We also inspected candidates from each contextual-keyword stratum, including occurrences with no recorded model-related context.

The extractor cannot recognize a genuinely new root or family absent from the dictionary. Severe OCR or text-extraction errors may obscure a known root, and unusual abbreviations or transliterations may be missed. Sentence segmentation is imperfect in corrupted text, tables, and reference lists. Conversely, the deliberately permissive candidate rules admit irrelevant occurrences of ambiguous names. Such false candidates are an expected product of this stage and motivate the separate model-identity classification described next.

\subsection{Model-Identity and Author-Use Classification}
\label{app:classification}

\subsubsection{Two classification problems}

Dictionary extraction identifies strings that might refer to known model families, but it does not by itself establish model use. We separate two questions. First, does the matched string actually denote the intended LLM family? Second, if it does, did the current article's authors use that model?

These questions produce different kinds of negative cases. An occurrence of \emph{SOLAR}, \emph{Orca}, or \emph{Gemini} may denote something other than the corresponding language model. Alternatively, the occurrence may genuinely identify the model but appear only in background discussion, related work, or a description of another study. We therefore trained separate model-identity and author-use classifiers.

The identity requirement was applied only to a predefined set of collision-prone families. For more distinctive families, dictionary recognition itself provided sufficient evidence of identity. The author-use classifier was applied to every candidate occurrence. These decisions were made at the occurrence level because different passages in the same paper can play different roles.

\subsubsection{Shared supervision strategy}

We developed the classification tasks through manual coding before using an LLM to produce training data. For each task, the authors first defined the relevant categories, resolved recurrent boundary cases, and constructed a gold-standard set of manually coded passages. We then translated those definitions into detailed annotation prompts and evaluated whether DeepSeek-V4-Flash could reproduce the human judgments. Prompts are given by Section~\ref{app:prompts}.

Once its agreement was adequate for the distinction required by the analysis, DeepSeek labeled a larger purpose-built sample. These labels constitute a \emph{silver standard}: they scale human-developed coding rules but remain model-generated rather than authoritative ground truth. Finally, we fine-tuned SciBERT 
 on the silver labels for corpus-scale inference.

The human and cross-validation results consequently evaluate different stages of the process. Comparisons between DeepSeek and the authors' gold-standard annotations assess the quality of the silver-label generator. SciBERT cross-validation assesses whether the smaller classifier reproduces those silver labels on held-out passages. We report these evaluations separately.

\subsubsection{Model-identity classification}

\paragraph{Human coding scheme and gold standard.}

The model-identity task asks whether a candidate occurrence denotes the model family proposed by the dictionary. An occurrence is \texttt{true\_llm} when it refers to the intended family and \texttt{not\_llm} when it refers to an ordinary word, person, animal, gene, dataset, instrument, unrelated algorithm, or another technical system. Identity is distinct from author use: a genuine model reference in related work is positive for identity even though it is not use by the current paper's authors.

A researcher manually coded a stratified sample drawn from the collision-prone candidates. One indeterminate case was excluded and replaced with another coded passage, producing a 150-passage binary gold standard containing 99 genuine LLM references and 51 non-LLM collisions.

\paragraph{Annotation sample and prompt.}

We constructed a larger sample of 1,500 occurrences from 1,500 distinct papers. The sample targeted families with substantial collision risks and was balanced approximately evenly across sampled families. Within families, we varied the contextual conditions recorded during extraction: explicit LLM language in the matched sentence, explicit LLM language elsewhere in the paper only, model-operation or NLP-task language, transformer terminology alone, provider-qualified or otherwise distinctive aliases, and no recorded contextual evidence. We also varied the matched surface---bare names, numbered releases, parameter-sized releases, and variant-labeled surfaces---and whether another model family appeared nearby.

The final sample contained 421 passages with no current contextual evidence, 301 with explicit LLM terminology elsewhere in the paper only, 161 with a provider-qualified or otherwise distinctive alias, 285 with explicit LLM terminology in the matched sentence, 241 with model-operation or NLP-task language, and 91 with transformer terminology alone. At least one additional (beyond the target) recognized family appeared in 729 passages and did not appear in 771. This design deliberately included difficult collisions rather than approximating the more positive distribution of the full corpus.

We translated the human coding distinction into a prompt asking whether a marked expression denoted the proposed LLM family. For this task, DeepSeek saw the naturally occurring target name enclosed in \texttt{<TARGET>} tags. We did not hide the name because distinguishing, for example, SOLAR the language model from ordinary solar terminology is the identity task itself. Other model names also remained visible. The prompt explicitly instructed DeepSeek not to consider whether the current article's authors used the model.

DeepSeek could assign \path{true_llm}, \path{not_llm}, or \path{uncertain}. Against the 150 author-coded binary cases, it achieved 0.951 precision, 0.990 recall, and 0.970 F1 for genuine LLM identity, with 0.953 overall accuracy.

\paragraph{Silver standard.}

DeepSeek labeled 968 of the 1,500 sampled passages \texttt{true\_llm}, 524 \texttt{not\_llm}, and eight \texttt{uncertain}. The sample covered the 41 families initially designated as collision-prone.

\paragraph{SciBERT training and evaluation.}

For SciBERT, the target occurrence was replaced with \texttt{[TARGET\_MODEL]}, while other model names remained visible. SciBERT therefore could not memorize which target families were especially collision-prone and had to learn contextual evidence of model identity.

Evaluation used five-fold cross-validation. Because the sample contained one occurrence from each paper, no paper could appear in both the training and held-out portions of a fold. We selected the training epoch using out-of-fold macro-F1. The third epoch performed best, yielding 0.921 accuracy, 0.911 macro-F1, and 0.941 F1 for genuine LLM identity. Genuine-reference precision was 0.914 and recall was 0.969.

Each fold initialized {\def\UrlFont{\ttfamily}\path{allenai/scibert_scivocab_uncased}} and trained for up to three epochs with AdamW, learning rate $2\times10^{-5}$, weight decay 0.01, 10\% linear warmup, training batch size 16, evaluation batch size 128, balanced class weights, and a maximum input length of 512 wordpieces. The final deployed classifier was trained for the selected three epochs on all 1,500 silver-standard training passages. Training and fold assignment used seed 13.

For diagnostic purposes, we scored all canonical candidate occurrences with the identity classifier. In the final decision rule, however, a positive identity prediction was required only for the predefined collision-prone families. Requiring classifier confirmation for every family would expose distinctive model names to unnecessary false-negative error.

\subsubsection{Author-use classification}

\paragraph{Development of the coding scheme.}

The second question is whether the current article's authors used the model. We initially developed a three-way coding scheme because the negative cases posed two substantively different attribution problems:

\begin{enumerate}
    \item \texttt{author\_use}: the current article's authors used, evaluated, queried, fine-tuned, trained, compared, prompted, embedded with, annotated with, generated data with, or otherwise operationally relied on the target model;
    \item \texttt{mention\_only}: the target appeared as background, context, motivation, an example, a general capability claim, future work, a limitation, or a comparison point without evidence of operational use; and
    \item \texttt{use\_by\_other}: the target was used in a cited study, benchmark, report, model card, or other external work, but not by the current article's authors.
\end{enumerate}

The distinction between \path{mention_only} and \path{use_by_other} was useful during manual coding and prompt development. Detailed results from a cited study can closely resemble results reported by the current authors unless the coder attends carefully to attribution. Retaining a separate \path{use_by_other} category allowed us to make that failure mode explicit in the annotation instructions.

For example, ``we evaluated \texttt{[TARGET\_MODEL]} on three benchmarks'' is \texttt{author\_use}. ``Prior work evaluated \texttt{[TARGET\_MODEL]}'' is \texttt{use\_by\_other}, while a general statement that the model exists or has particular capabilities is \texttt{mention\_only}.

A researcher manually coded 150 passages using this three-way scheme. The gold standard contained 111 author-use cases, 33 mention-only cases, and six uses attributed to others.

\paragraph{Prompt construction and validation.}

We translated the manual coding scheme into a prompt whose central question was: ``Did the authors of the current article mention \texttt{[TARGET\_MODEL]} because they used it in their own study, or did they merely mention it?'' The prompt emphasized that ``the authors'' meant the authors of the current article, not the authors of a cited paper, benchmark, model card, software package, or external report.

Each prompt supplied a passage together with its section label and title. The target name was replaced with \texttt{[TARGET\_MODEL]}, while other model names remained visible. Masking prevented the annotator from deciding that a model was likely to be used merely because of the identity or popularity of its family. The prompt instructed DeepSeek to use section metadata only as a prior, classify only the target occurrence, give a short evidence-based rationale, and return exactly one of the three labels in structured form. The complete prompt and user-message template are included in the replication archive.

Against the 150 manually coded passages, DeepSeek achieved 0.880 three-class accuracy. Its F1 scores were 0.943 for \texttt{author\_use}, 0.767 for \texttt{mention\_only}, and 0.471 for \texttt{use\_by\_other}. The main weakness was therefore not recognizing author use, but separating the two ways in which an occurrence could be negative.

The distinction between those two negative classes was conceptually useful but unnecessary for the paper's analytic outcome. Both indicate that the current article's authors did not use the model. We therefore retained the three-way task in the prompt and silver annotation, but consolidated \path{mention_only} and \path{use_by_other} before training SciBERT. Under this binary coding, DeepSeek achieved 0.916 accuracy and 0.943 author-use F1 against the human labels, with 0.950 precision and 0.935 recall.

\paragraph{Silver standard.}

DeepSeek labeled a purpose-built sample of 1,500 passages drawn from 205 papers. It assigned 1,110 passages to \path{author_use}, 272 to \path{mention_only}, and 118 to \path{use_by_other}. Consolidating the latter two categories produced the binary silver standard used for SciBERT: 1,110 author-use passages and 390 non-use passages.

Thus, the three-way distinction shaped the annotation prompt and remains available in the archived silver labels, but it was not the outcome learned by SciBERT. The deployed classifier distinguishes only author use from non-use.

\paragraph{SciBERT training and evaluation.}

We fine-tuned SciBERT on the 1,500 passages using the consolidated binary label. Its input included up to 205 tokens on either side of the target occurrence, subject to the 512-wordpiece sequence limit. As in the DeepSeek annotation stage, the target was replaced with \texttt{[TARGET\_MODEL]} and other model names remained visible.

Evaluation used five-fold cross-validation grouped by paper identifier. All occurrences from a paper were assigned to the same fold, preventing passages from one article from appearing in both training and validation data. Across folds, SciBERT achieved 0.89 accuracy and 0.93 F1 for the author-use class. A class-balanced logistic regression using unigram and bigram TF--IDF features achieved 0.86 accuracy and 0.90 author-use F1 under the same grouped folds.

\begingroup\sloppy
Cross-validation initialized \path{allenai/scibert_scivocab_uncased} in each fold and trained for three epochs with AdamW, learning rate $2\times10^{-5}$, weight decay 0.01, 10\% linear warmup, maximum gradient norm 1, training batch size 8, evaluation batch size 16, and balanced class weights. Inputs were truncated to 512 wordpieces. The final deployed model used the same settings and was fit for one epoch on all 1,500 passages. Training and fold assignment used seed 13. At inference, the predicted label was the class with the largest softmax probability; no probability threshold was tuned.
\par\endgroup

\subsubsection{Combined model-use decision rule}

The classifiers were combined at the occurrence level. A candidate occurrence satisfied the identity rule when either its family was not collision-prone or the identity classifier labeled it as a genuine LLM reference. In either case, the occurrence also had to be classified as author use.

Formally, for candidate occurrence $m$ with family $f(m)$, let $I_m$ indicate a positive model-identity prediction, $U_m$ indicate a positive author-use prediction, and $\mathcal{C}$ denote the set of collision-prone families. The occurrence qualified as model use when

\begin{equation}
U_m=1
\quad\text{and}\quad
\left[f(m)\notin\mathcal{C}\ \text{or}\ I_m=1\right].
\label{eq:model_use_rule}
\end{equation}

A paper--family pair was retained when at least one occurrence in the paper met this rule. Repeated qualifying occurrences and multiple releases belonging to the same family did not create additional paper--family records.

This rule establishes that the occurrence denotes the intended family, where confirmation was necessary, and that the current article's authors reported operational reliance on it. The model-identity rule removed 118,903 articles and 
406,351 model occurrences. Subsequently applying the author-use classifier removed 42,353 articles and 
846,783 occurrences. See Table~\ref{tab:sample_flow}.

The author-use classifier was not originally designed to distinguish use as a scientific instrument or object from ancillary research-workflow assistance, such as editing or manuscript preparation. In our original manual coding we did not encounter such cases in any notable frequency and only later came to understand the scale of the problem. We address this narrower boundary with the downstream disclosure-proxy classifier described in the following section.

\subsection{AI-Disclosure Statement Classification}
\label{app:disclosure}
In manually auditing model occurrences, we encountered statements about LLM-use as research workflow tools. Such uses were frequently found in AI-disclosure statements. A targeted audit of AI-disclosure statements revealed frequent reports about the use of tools aiding manuscript preparation, largely to improve grammar, flow, and formatting. However, we also observed statements relating other scientific workflow tools such as code generation, literature search, and research design consultation.

Accordingly, we mined the S2ORC full text for variants of AI-disclosure section headings. Manual review of 150 AI-disclosure statements found that 71\% reported manuscript preparation use and 21\% reported other workflow uses; 11\% reported no use of AI tools. We also found genuine occurrences of model use. 9\% of predicted disclosure statements reported the use of AI as scientific instruments or objects including classification, benchmarking, and model evaluation. These categories were not mutually exclusive, as a single model reference could refer to multiple uses.

Given that 91\% of the model occurrences in AI-disclosure statements did not reflect use as a scientific instrument or object, we trained a third, weakly supervised classifier to identify disclosure-like passages as a proxy for non-instrument use.

\subsubsection{Sample construction}
Positive cases were model-mention passages appearing under strict variants of AI-disclosure section headings. We retained all such passages regardless of their upstream author-use classification. Thus, the positive class intentionally included statements reporting manuscript assistance, other workflow assistance, and no use, as well as some genuine scientific-instrument or object uses. This procedure produced 4,757 unique positive passages from 4,457 papers.

To ensure that the negative class represented both the deployment population and the scientifically consequential boundary cases, we sampled equal numbers of within-paper controls, clear scientific-instrument or object uses, and random qualifying occurrences. All negative occurrences had passed the upstream model-identity and author-use classifiers. The resulting silver standard contained 19,028 passages from 13,349 papers.

\subsubsection{SciBERT training and evaluation}
Section headings were used only to construct the silver labels and were excluded from classifier inputs. We also replaced every recognized model-namesurface in a passage with \texttt{[TARGET\_MODEL]}, preventing the classifier from learning that particular families, such as ChatGPT, were intrinsically associated with disclosure statements.

We then fine-tuned {\def\UrlFont{\ttfamily}\path{allenai/scibert_scivocab_uncased}} on the silver standard. Cross-validation used five folds grouped by paper, ensuring that no paper contributed passages to both the training and held-out portions of a fold. The folds were greedily balanced by class label. Each fold contained approximately 15,222 training passages and 3,806 held-out passages.

SciBERT was trained for three epochs per fold using AdamW, learning rate $2\times10^{-5}$, weight decay 0.01, 10\% linear warmup, maximum gradient norm 1, training batch size 16, evaluation batch size 64, and balanced class weights.
For model comparison, predictions were generated from the epoch with the largest held-out macro-F1 in each fold. We report out-of-fold precision, recall, and F1 for the disclosure-proxy class, macro-F1, and false-positive rates separately for the three negative strata.

SciBERT achieved 0.982 accuracy, 0.976 macro-F1, and 0.964 F1 for the disclosure-proxy class.

Our primary concern was avoiding the erroneous removal of genuine scientific-instrument or object uses. We therefore applied the final classifier using a conservative disclosure probability threshold of 0.95, rather than the conventional 0.50 decision threshold. At this threshold, the classifier retained 94.7\% of the 4,757 strict-heading positives. It incorrectly flagged 53 of 4,757 within-paper negative controls (1.11\%), 49 of 4,757 randomly sampled qualifying occurrences (1.03\%), and only 3 of 4,757 clear scientific-instrument or object uses (0.06\%). Across all three negative strata, the false-positive rate was 0.74\%.

After cross-validation, we trained a final SciBERT model on all 19,028 silver-standard passages and applied it to the 1,542,933 passages containing the 2,366,252 occurrences that had passed the model-identity and author-use stages. At the 0.95 threshold, the classifier identified 16,333 disclosure-like passages, corresponding to 19,566 qualifying occurrences in 15,287 papers. Removing these occurrences eliminated 14,557 paper--family pairs. A paper was removed only when every otherwise qualifying occurrence was classified as disclosure-like. Consequently, 11,080 papers lost all qualifying model families, reducing the full-corpus sample from 178,531 to 167,451 papers. 72.3\% of the removed family--paper pairs were due to ChatGPT mentions; the leading five families---ChatGPT, GPT, Claude, Gemini, and DeepSeek---account for account for 98.5\% of all removed pairs. The final January 2023--June 2026 analytic sample contained 157,446 papers.

This classifier should be interpreted as a proxy for disclosure-like language, rather than as an exhaustive classification of research-workflow use. Its weak labels deliberately capture the distinctive form and location of AI-disclosure statements, including some statements reporting no AI use and a small number describing genuine scientific uses. Conversely, workflow assistance reported outside disclosure-like passages may remain in the analytic sample. The high deployment threshold reflects the asymmetric cost of these errors: retaining some workflow disclosures adds noise, whereas excluding genuine scientific uses removes observations that belong in the study.

\subsection{Statistical Models}
\label{app:models}
All statistical modeling was conducted in R. Packages and settings are specified in the text.

\subsubsection{Binary open-weight model}

For single-family paper $i$, let $Y_i=1$ when its sole family is open-weight and $Y_i=0$ otherwise. The principal logistic specification is
\begin{equation}
\operatorname{logit}\{\Pr(Y_i=1)\}
= \alpha + \mathbf{x}_i^\top\boldsymbol{\beta}
+ f(t_i) + u_{s[i]},
\label{eq:binary}
\end{equation}
where $\mathbf{x}_i$ contains the mutually exclusive China-link indicators, $f(t_i)$ is a natural spline in exact publication date, and $u_{s[i]}$ is a random intercept for OpenAlex subfield $s$. Model~1 omits $u_{s[i]}$ and includes Chinese institutional affiliation; Model~2 adds subfield random intercepts; and Model~3 additionally distinguishes Chinese-origin names outside China.

Publication time was decimal year. The date adjustment was an R \texttt{splines::ns} natural spline with four degrees of freedom; its boundary and interior knots were selected by the function from the observed dates rather than fixed manually.  For records dated January 1 because only a publication year was known (4,001 papers, 8\% of the sample), the spline basis was averaged over the four quarter-start dates for that year.

The mixed models were estimated with \texttt{lme4::glmer}, a logit link, normally distributed subfield random intercepts, and the \texttt{bobyqa} optimizer with \texttt{maxfun=100000}. Reported 95\% confidence intervals are Wald intervals, $\exp(\hat\beta\pm1.96,\mathrm{SE})$. The final 48,129-paper model had 83 subfield levels and converged without a recorded optimizer warning.

\subsubsection{Multinomial family-category model}

The complementary outcome has three categories: proprietary, other open-weight, and China-developed open-weight. Proprietary is the reference category. For category $k$ in the two open-weight categories,
\begin{equation}
\log\frac{\Pr(Y_i=k)}{\Pr(Y_i=\text{proprietary})}
=\alpha_k+\mathbf{x}_i^\top\boldsymbol{\beta}_k+f_k(t_i)
+\boldsymbol{\gamma}_{k,s[i]}.
\label{eq:multinomial}
\end{equation}
Here $\boldsymbol{\gamma}_{k,s[i]}$ denotes subfield fixed effects. Fixed effects serve the same adjustment goal as the subfield term in Equation~\ref{eq:binary} while keeping the multinomial model directly estimable. The model was fit with \texttt{nnet::multinom} on 48,125 complete cases, using the proprietary category as the outcome reference, \texttt{maxit=120}, and \texttt{MaxNWts=10000}. It included the same four-degree-of-freedom natural spline and indicators for all observed primary subfields (230 levels, with the software-selected treatment-coded reference). Unlike the binary random-intercept model, these fixed effects do not partially pool small subfields; their role here is adjustment rather than reporting field-specific estimates.

Reported quantities are average adjusted probabilities, standardized over the pooled distribution of dates and subfields. We compute uncertainty by drawing jointly from the estimated coefficient covariance matrix and recomputing all adjusted probabilities and contrasts in each draw. We used 500 multivariate-normal coefficient draws (seed 20260710) and report the 2.5th and 97.5th percentiles. January 1st dates were randomly distributed across the known year as in the binary model.

\subsubsection{Decomposition of the adoption increase}

Let $p_{gt}$ be open-weight adoption in group $g$ and year $t$, and let $w_{gt}$ be the group's sample share. Aggregate adoption is $P_t=\sum_g w_{gt}p_{gt}$. We decompose the change between 2023 and the June 2026 partial-year endpoint using the symmetric two-period decomposition
\begin{align}
P_{2026}-P_{2023}
={}&\sum_g \bar{w}_g(p_{g,2026}-p_{g,2023}) \nonumber\\
&+\sum_g \bar{p}_g(w_{g,2026}-w_{g,2023}),
\label{eq:decomposition}
\end{align}
where $\bar{w}_g=(w_{g,2026}+w_{g,2023})/2$ and $\bar{p}_g=(p_{g,2026}+p_{g,2023})/2$. This is the implemented allocation rule. Uncertainty for each total contribution was obtained from 2,000 bootstrap resamples within publication year (seed 20260630), using the 2.5th and 97.5th percentiles.

\begin{figure}[t]
\centering
\includegraphics[width=\linewidth]{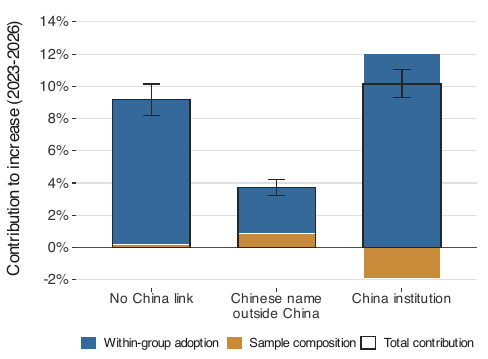}
\caption{Decomposition of single-family-paper open-weight model adoption by Chinese-linked authorship, 2023--June 2026, into within-group adoption and sample-composition components.}
\label{fig:chinese_link_open_adoption_decomposition}
\end{figure}

The results of this analysis, described in the main paper, are visualized in Figure~\ref{fig:chinese_link_open_adoption_decomposition}.

\subsection{AI Assistance and Reproducibility}
\label{app:reproducibility}

\subsubsection{Generative-AI assistance and author responsibility}

Codex (GPT-5.5 \& 5.6) was used for model dictionary construction, Chinese-name dictionary construction, code development and debugging, and generally aided in designing components of the analytical strategy. Codex was also employed heavily in drafting these appendices; all material was reviewed and edited by the authors. The substantive uses of another LLM as an annotator are reported separately in Section~\ref{app:classification}. The authors are responsible for the final taxonomy, analysis, interpretation, and manuscript text.

\subsubsection{Reproducibility}
The public \href{https://github.com/zackarydunivin/who-uses-open-weights}{reproducibility repository} contains the model dictionary for extraction; all 3,619,386 extracted candidate occurrences in a compressed, text-free relational format; LLM prompts, structured annotations, manually classified evaluation data, and fixed training and validation assignments; extraction, classifier-training, inference, and analysis code; software specifications, random seeds, and machine-readable values underlying the reported tables and figures. A streamlined workflow reproduces the statistical results from frozen classification decisions and automatically checks the resulting sample counts, tables, figures, and model outputs against the reported values.

An optional end-to-end workflow starts from independently obtained S2ORC shards and rebuilds extraction, canonical-source selection, all three classifiers, inference, and the final analytic cohort. The three fine-tuned SciBERT checkpoints are distributed as versioned release assets with checksums and model cards. The repository does not redistribute complete articles, S2ORC shards, or raw bibliographic API responses; it provides the exact SciBERT base-model revision, document and shard identifiers, within-document locations, and the short S2ORC-derived training and audit windows needed to reconstruct and examine the measurement process.

\subsection{LLM Annotation Prompts}
\label{app:prompts}

This section reproduces the system prompts and user-message templates used to generate the silver-standard annotations. Both tasks used DeepSeek-V4-Flash with temperature 0, top-$p=1$, and reasoning mode disabled. Braced expressions in the user-message templates indicate fields populated separately for each occurrence. Visual line wrapping below does not represent additional prompt content.

\subsubsection{Model-identity annotation}

For model identity, the observed target name remained visible and was enclosed in \texttt{<TARGET>} tags. Other model names appeared as written in the source passage.

\paragraph{System prompt.}\leavevmode\par\nobreak

\begin{Prompt}
You are a careful entity-linking coder for a study of large language models in scientific articles.

You will receive one scientific-text passage. The occurrence to judge is enclosed in <TARGET>...</TARGET>. All other names in the passage are shown exactly as written. The observed target surface and the intended LLM family are also shown separately.

Your only task is to decide whether the marked target occurrence refers to the intended large-language-model family. Do not decide whether the article's authors used the model; background mentions and use by other researchers are still true LLM references.

Labels:

1. true_llm

Choose true_llm when the marked occurrence denotes the intended LLM, language-model family, release, or variant. This includes mentions in background, comparisons, citations, and descriptions of another study.

2. not_llm

Choose not_llm when the marked occurrence has another sense. Examples include a person, animal, plant, gene or protein, medical intervention, material, instrument, satellite, algorithm, statistical model, dataset, software package, mathematical symbol, ordinary word, or an unrelated acronym.

3. uncertain

Choose uncertain only when the supplied passage genuinely does not contain enough information to distinguish the intended LLM from another plausible sense. Do not use uncertain merely because the mention is brief.

Important safeguards:

- Judge only the occurrence enclosed in <TARGET>...</TARGET>.
- Other model names may provide legitimate context, but their presence does not by itself prove that the marked target is the intended LLM.
- The observed target surface is legitimate evidence. Provider-qualified names, model-specific variants, release notation, parameter sizes, and explicit language-model wording may support true_llm.
- Generic machine-learning language alone is not sufficient if the target clearly names a different scientific or technical entity.
- A numeral or generic suffix does not automatically turn a collision-prone word into an LLM name.

Give a short rationale, then exactly one JSON label.

Format:

**Rationale:**
[One or two sentences]

**Label:**
```json
{
  "label": "true_llm"
}
```

The only acceptable labels are true_llm, not_llm, and uncertain. Do not write anything after the JSON.
\end{Prompt}

\paragraph{User-message template.}\leavevmode\par\nobreak

\begin{Prompt}
Target metadata:

observed_surface: {target_matched_text}
intended_llm_family_id: {target_family_id}
intended_llm_family_name: {target_family_name}
intended_provider: {target_provider}

Article metadata:

section_label: {section_label}
section_title: {section_title}

Passage:

{target_marked_text}

Code the marked target occurrence.
\end{Prompt}

\subsubsection{Author-use annotation}

For author use, the target occurrence was replaced with \texttt{[TARGET\_MODEL]}. Other model names remained visible. The prompt retained the original three-way coding scheme even though \texttt{mention\_only} and \texttt{use\_by\_other} were subsequently combined into a single non-use class for SciBERT training.

\paragraph{System prompt.}\leavevmode\par\nobreak
\begin{Prompt}
You are a careful qualitative coder for a study of how scientific articles mention and use large language models.

Task description:
You will receive one snippet from a scientific article. In the snippet, one model mention has been replaced with [TARGET_MODEL]. Your task is to code only that target mention.

The central question is:
Did the authors of the current article mention [TARGET_MODEL] because they used it in their own study, or did they merely mention it?

Use the section metadata as a prior, but let the text decide. Methods, Results, Discussion, and Abstract snippets often describe the current article's own work. Introduction, Related Work, and Background snippets often describe prior work or general context. Unknown sections may be any of these, so read the section title and snippet carefully.

Important distinction:

"The authors" means the authors of the current scientific article containing this snippet. It does not mean authors of a cited paper, survey, benchmark, model card, software package, or external report mentioned inside the snippet.

Primary labels:

1. author_use

Choose author_use when the current article's authors used, evaluated, queried, fine-tuned, trained, compared, prompted, embedded with, annotated with, generated data with, or otherwise operationally relied on [TARGET_MODEL] as part of their own study.

2. mention_only

Choose mention_only when [TARGET_MODEL] is mentioned as background, context, motivation, an example of a model, general adoption statistics, broad field description, future work, limitations, or a comparison point with no clear evidence that the current article's authors used it.

This includes cases where:
- The snippet says [TARGET_MODEL] exists or is popular.
- The snippet describes general capabilities, risks, or social context of [TARGET_MODEL].

3. use_by_other

Choose use_by_other when [TARGET_MODEL] was used in a cited or external study, but not by the current article's authors. This label is for prior-work empirical results that can look like current-paper results unless you attend to attribution.

This includes cases where:
- The snippet attributes use or results to named external authors, a citation, a previous study, a prior benchmark, a report, or a model provider.
- The snippet summarizes detailed empirical results from another paper.

Target specificity:

Code only the [TARGET_MODEL] placeholder. Other model names in the snippet may provide context, but they are not the target. If the snippet contains multiple model mentions, do not assign a label to the non-target models.

I will show you a snippet, then you will select the best label. Please explain why you made your decision with a short rationale.

Format your response as follows:

**Rationale:**
[Your reasoning for your decision]

**Label:**
```json
{
  "label": "author_use"
}
```

Remember you must make a decision, even if the evidence is weak. The only acceptable responses for Label are author_use, mention_only, or use_by_other.

Do not write anything after "**Label:**" except the json.
\end{Prompt}

\paragraph{User-message template.}\leavevmode\par\nobreak

\begin{Prompt}
Section metadata:

section_label: {section_label}
section_title: {section_title}

Snippet:

{masked_text}

Code this record.
\end{Prompt}